\documentclass[3p,times]{elsarticle}

\usepackage{ecrc}

\volume{00}

\firstpage{1}

\journalname{Physics Letters B}

\runauth{Rajdeep Basak and Krishnendu Mukherjee}

\jid{plb}

\jnltitlelogo{Physics Letters B}

\usepackage{amssymb}
\usepackage{amsthm}

\usepackage[figuresright]{rotating}

\usepackage{epstopdf}
\usepackage{graphics}
\usepackage{graphicx}

\begin{document}

\begin{frontmatter}



\dochead{}

\title{Stable, finite density solutions in the effective theory of 
non-abelian gauge fields}


\author{Rajdeep Basak}
\author{Krishnendu Mukherjee}
\address{Department of Physics, Indian Institute of Engineering Science and
Technology, Shibpur, Howrah-711103\\
}

\begin{abstract}
We consider the gauge fixed partition function of pure
$SU(N_c)$ gauge theory in axial gauge following the
field strength formalism. We integrate over
$3 (N_c^2-1)$ field strengths using the Bianchi identities
and obtain an effective action of the remaining $3 (N_c^2-1)$
field strengths in momentum space. We obtain the static solutions of
the equations of motion (EOM) of the effective theory. The
solutions exhibit Gaussian nature in the $z$ component of
momentum and are proportional to delta functions of the
remaining components of momentum. 
The solutions render a stable, finite energy density 
in the strong coipling region and the parameters of the solution
are proportional to fourth root of the gluon condensate, which
acts as a natural mass scale in the low energy phase of the theory.

\end{abstract}

\begin{keyword}
Non-abelian gauge theory, Quantum Chromodynamics, Semi-nonperturbative QCD
\end{keyword}

\end{frontmatter}


\section{Introduction}

It is well known that non abelian gauge theory (NAGT) has asymptotic freedom
\cite{Gross1973} which indicates that the theory is almost free at very high
energy and at very small distance scale.
It also indicates that at very low energy
it exhibits the confinement of gauge degrees of freedom
and results in the non-zero values of
gluon condensates at this low energy phase of the theory\cite{Shifman1979}.
Owing to non linear nature of NAGT its exact quantization
is still lacking and as a consequence of which the dynamics
of this theory at low energy remains least understood till today.
It is believed that stable classical solutions of the 
equations of motion(EOM) of the theory may be
of use to understand the behaviour of the system at low energy
and it may also shed some light on the issue of quantization
of the theory at this energy\cite{Jackiw1977}.

The endeavour for obtaining the classical solutions of EOM
of NAGT has a long history since its inception.
An attempt has been made to caste those equations
in the form of Maxwell equations of electrodynamics\cite{Maciejko1978}.
The exact periodic solutions of the $SU(2)$
gauge theory have been constructed
in Minkowski space-time\cite{Oh1979}.
There also exist solutions which are the
non-abelian analogues of electromagnetic plane waves\cite{Coleman1977}.
A relationship has been established
between the solutions of a $\phi^4$ scalar field theory and a class
of solutions of EOM of $SU(2)$ gauge theory\cite{Corrigan1977}.
The most general self-dual, non-abelian, plane wave solutions
have also been obtained in NAGT\cite{Lo1980}.
A comprehensive discussion on the solutions of EOM
in Minkowski space-time are given in
Ref.\cite{Actor1979, Malec1987}

It has been proposed long before\cite{Halpern1979} that a non-abelian gauge
theory can be formulated in terms of field strengths in axial gauge.
The unique inversion of gauge potentials to field strengths requires
Bianchi identities to be satisfied in the quantized theory. We use
Bianchi identities to integrate over the $3(N_c^2-1)$ degrees of
freedom and obtain an effective action in terms of the remaining
$3(N_c^2-1)$ field strengths in the momentum space.
Then we obtain a solution of the equations of motion of
the effective theory. Solutions indicate that the
non abelian electric and magnetic fields have Gaussian nature in their $z$
component of the momentum.
We expand the effective action around this solution to quadratic
order in fluctuation and obtain the energy density of the system
by integrating out the fluctuations in the partition function.
The energy density has been found to have minima with respect
to one of the parameters of the solution, which indicates
the stability of the solutions under the variation of the parameter.

\section{Gauge Fixed Effective Action}

We start with the pure $SU(N_c)$ gauge theory
Lagrangian
\begin{equation}
{\cal{L}} = -\frac{1}{4}G^{a\mu\nu}G_{\mu\nu}^a
\end{equation}
where $G^a _{\mu\nu}=\partial_{\mu}A^a _{\nu}-\partial_{\nu}A^a _{\mu}
+gC^{abc}A^b _{\mu}A^c _{\nu}$ ($a\,,b\,,c = 1,\,\cdots,\,N_c^2-1$)
is the field strength of the non-abelian
gauge field $A^a_\mu$. Here the axial gauge $A_3^a(x)=0$ is chosen.
The unique inversion $A(G)$ makes it possible to
change variables to field strengths and the gauge fixed
partition function takes the form\cite{Halpern1979}
\begin{equation}
Z=\int\mathcal{D}G\delta[I[G]]\exp
\Big[ -\frac{i}{4}\int d^4x G^a_{\mu\nu}G^{a\mu\nu}\Big].
\label{Zx}
\end{equation}
where
\begin{equation}
\delta[I[G]]=\prod_{\mu, a,x}\delta(I^a_{\mu}(x)),
\end{equation}
\begin{equation}
I^{a\mu}(x) = \partial_\lambda\widetilde{G}^{a\lambda\mu}
+gC^{abc}A_\lambda^b(G)\widetilde{G}^{c\lambda\mu}
\end{equation}
and
\begin{equation}
\widetilde{G}^{a\mu\nu} =
\frac{1}{2}\epsilon^{\mu\nu\lambda\sigma}G^a_{\lambda\sigma}.
\label{IamuGx}
\end{equation}
We use the Fourier transform of
\begin{equation}
G^a_{\mu\nu}(x) = \int\frac{d^4k}{(2\pi)^4} e^{-ik.x} G^a_{\mu\nu}(k)
\end{equation}
and
\begin{equation}
A^a_\mu(x) = \int\frac{d^4k}{(2\pi)^4} e^{-ik.x} A^a_{\mu}(k)
\label{FTGA}
\end{equation}
and write the solutions of Halpern\cite{Halpern1979} for the gauge
potentials in momentum space as
\begin{equation}
A^a_0(k) = -\frac{i}{k_z}E^a_z(k),~~
A^a_1(k) = -\frac{i}{k_z}B^a_y(k),~~
A^a_2(k) = \frac{i}{k_z}B^a_x(k),~~
A^a_3(k)=0.
\label{solA}
\end{equation}
${\mathbf E}^a(k)$ and ${\mathbf B}^b(k)$ are the non abelian
electric and magnetic field vectors in momentum space. We use
eq.(\ref{FTGA}) and (\ref{solA}) to write the partition function in
terms of the fields in momentum space as
\begin{equation}
Z=\int\mathcal{D}G\delta[I[G]] e^{iS[G]},
\label{Zk}
\end{equation}
where
\begin{equation}
S[G] = -\frac{1}{4}\int\frac{d^4k}{(2\pi)^4} 
G^a_{\mu\nu}(k)G^{a\mu\nu}(-k)\label{SG}
\end{equation}
and
\begin{equation}
\delta[I[G]] = \prod_{\mu, a,k}\delta(I^a_{\mu}(k)).
\end{equation}
$I^{a\mu}(k)$ in momentum space in terms of non abelian
electric and magnetic fields read
\begin{eqnarray}
I^{a0}(k) &=& -i{\mathbf k}.{\mathbf B}^a(k)
+igC^{abc}\int\frac{d^4k^\prime}{(2\pi)^4}\frac{1}{k_z^\prime}
({\mathbf B}^b(k-k^\prime){\mathbf\times}{\mathbf B}^c(k^\prime))_z,
\label{Ia0}\\
{\mathbf I}^a(k) &=& -ik_0{\mathbf B}^a(k)
+i{\mathbf k}{\mathbf\times}{\mathbf E}^a(k)
+igC^{abc}\int\frac{d^4k^\prime}{(2\pi)^4}\frac{1}{k_z^\prime}
[-{\mathbf B}^b(k-k^\prime)E^c_z(k^\prime)
+E^b_z(k-k^\prime)(\hat{x}B^c_x(k^\prime)+\hat{y}B^c_y(k^\prime))\nonumber\\
& &-\hat{z}(E^b_x(k-k^\prime)B^c_x(k^\prime)
+E^b_y(k-k^\prime)B^c_y(k^\prime))],
\label{Iavec}
\end{eqnarray}
where $\hat{x}$, $\hat{y}$, $\hat{z}$ are the unit vectors along
$x$, $y$ and $z$ axes respectively. We write
\begin{equation}
I^{a0}(k) = -ik_z(B^a_z(k)-f^a_3(k)),~~
I^a_x(k) = -ik_z(E^a_y(k)-f^a_2(k)),~~
I^a_y(k) = ik_z(E^a_x(k)-f^a_1(k)).
\label{I0xy}
\end{equation}
$f^a_j(k)$ in terms of the fields $X^a_j(k)$ ($j=1,2,3$)
are given as
\begin{eqnarray}
f^a_1(k)&=&\frac{k_xX^a_3(k)+k_0X^a_2(k)}{k_z}
-gC^{abc}\int\frac{d^4k'}{(2\pi)^4}
\frac{X^b_2(k-k')X^c_3(k')}{k'_z(k'_z-k_z)},\label{f1}\\
f^a_2(k)&=&\frac{k_yX^a_3(k)-k_0X^a_1(k)}{k_z}
-gC^{abc}\int\frac{d^4k'}{(2\pi)^4}
\frac{X^b_3(k-k')X^c_1(k')}{k'_z(k'_z-k_z)},\label{f2}\\
f^a_3(k)&=&\frac{-(k_xX^a_1(k)+k_yX^a_2(k))}{k_z}
-gC^{abc}\int\frac{d^4k'}{(2\pi)^4}
\frac{X^b_1(k-k')X^c_2(k')}{k'_z(k'_z-k_z)},\label{f3}
\end{eqnarray}
where we have defined
\begin{equation}
X^a_1=B^a_x,~~X^a_2=B^a_y,~~X^a_3=E^a_z.
\end{equation}
We write using eq.(\ref{I0xy})
\begin{equation}
\delta[I(G)] = \prod_{a,k}\frac{i}{k_z^3}\delta(E^a_x(k)-f^a_1(k)) 
\delta(E^a_y(k)-f^a_2(k))
\delta(B^a_z(k)-f^a_3(k))\delta(I^a_z(k))
\end{equation}
and then performing integration over the fields $E^a_x$, $E^a_y$ and 
$B^a_z$, we obtain
\begin{equation}
Z={\cal N}\int\mathcal{D}X e^{iS [X]},
\label{Z}
\end{equation}
where ${\cal N}=\delta(0)\prod_{k_z}\frac{i}{k_z^3}$.
The factor of $\delta(0)$ comes from the factor $\delta(I^a_z(G))$ in the 
partition function when $I^a_z(G)$ becomes zero after the integration 
over the fields have been performed. The effective action
\begin{equation}
S[X]=S^{(0)}[X]+S^{(1)}[X],
\label{SX}
\end{equation}
where the quadratic part
\begin{equation}
S^{(0)}[X]=\frac{1}{2}\int \frac{d^4p}{(2\pi)^4}
X^a_j(p)(M^{-1})_{jk}(p)X^a_k(-p)
\end{equation}
and the interaction part
\begin{eqnarray}
S^{(1)}[X]&=&gC^{abc} \int
\frac{d^4p}{(2\pi)^4}\frac{d^4q}{(2\pi)^4}
\frac{b^{(3)} _{jkl}(p)}{p_zq_z(p_z-q_z)}
X^a_j(-p)X^b_k(p-q)X^c_l(q)\nonumber\\
& &+\frac{1}{2}g^2C^{abc}C^{ade}\int 
\frac{d^4p}{(2\pi)^4}\frac{d^4q}{(2\pi)^4}\frac{d^4q'}{(2\pi)^4}
\frac{b^{(4)} _{jklm}}{q_zq'_z(q_z-p_z)(q'_z+p_z)}
X^b_j(p-q)X^c_k(q)X^d_l(-p-q')X^e_m(q').
\end{eqnarray}
Here $3\times 3$ matrix $M^{-1}(p)$ reads
\begin{equation}
M^{-1}(p)=
\left( \begin{array}{ccc}
\frac{p^2+p_y^2}{p_z^2} & -\frac{p_xp_y}{p_z^2} 
& -\frac{p_0p_y}{p_z^2}\\
-\frac{p_xp_y}{p_z^2} & \frac{p^2+p_x^2}{p_z^2}
& \frac{p_0p_x}{p_z^2}\\
-\frac{p_0p_y}{p_z^2} & \frac{p_0p_x}{p_z^2} 
& \frac{|{\bf{p}}|^2}{p_z^2}\\
\end{array} \right),
\end{equation}
\begin{eqnarray}
b^{(3)}_{jkl}(p)&=& (p_x\delta_{j3}
+p_0\delta_{j2})\delta_{k2}\delta_{l3}
+(p_y\delta_{j3}-p_0\delta_{j1})\delta_{k3}\delta_{l1}
+(p_x\delta_{j1}+p_y\delta_{j2})\delta_{k1}\delta_{l2},\\
b^{(4)}_{jklm}&=&\delta_{j2}\delta_{k3}\delta_{l2}\delta_{m3}
+\delta_{j3}\delta_{k1}\delta_{l3}\delta_{m1}
-\delta_{j1}\delta_{k2}\delta_{l1}\delta_{m2}.
\end{eqnarray}

\section{Solution of EOM}

Equations of motion
\begin{equation}
\Big(\frac{\delta S[X]}{\delta X^a_j(p)}\Big)_{X=\bar{X}} =0
\end{equation}
take the form
\begin{eqnarray}
&&(M^{-1})_{kl}(p)\bar{X}^a_l(-p)
-gC^{abc}\frac{1}{p_z}
\int \frac{d^4q'}{(2\pi)^4}
\frac{\tilde{b}^{(3)}_{klm}(p,q')}{q'_z(p_z+q'_z)}
\bar{X}^b_l(-p-q')\bar{X}^c_m(q')\nonumber\\
& &-\frac{1}{2}g^2C^{abe}C^{ecd}\frac{1}{p_z}
\int \frac{d^4q'}{(2\pi)^4}\frac{d^4q''}{(2\pi)^4}
\frac{\tilde{b}^{(4)}_{klmn}}{q'_zq''_z(p_z+q'_z+q''_z)}
\bar{X}^b_l(q')\bar{X}^c_m(-q'-q''-p)\bar{X}^d_n(q'') =0,
\label{eom1}
\end{eqnarray}
where
\begin{eqnarray}
\tilde{b}^{(3)}_{klm}(p,q)&=&b^{(3)}_{klm}(p)+b^{(3)}_{lkm}(p+q)
-b^{(3)}_{mlk}(q),\\
\tilde{b}^{(4)}_{klmn}&=&b^{(4)}_{klmn}-b^{(4)}_{lkmn}+b^{(4)}_{mnkl}
-b^{(4)}_{mnlk}.
\end{eqnarray}
We assume a solution of the form
\begin{equation}
\bar{X}_j^a(p)=(2\pi)^4 \delta^{(3)}(p_{\perp})\xi^a_j(p_z),
\end{equation}
where $p_\perp=(p_0, p_x, p_y)$ and
\begin{equation}
\xi^a_j(z)\rightarrow 0
~~{\rm as}~~ |z|\rightarrow \infty. 
\label{propxi}
\end{equation}
Upon substitution of this solution
in eq.(\ref{eom1}), we obtain
\begin{eqnarray}
& &\{-\xi^a_1(-p_z)\delta_{n1}-\xi^a_2(-p_z)\delta_{n2}
+\xi^a_3(-p_z)\delta_{n3} \}\nonumber\\ 
& &-\frac{1}{2p_z}g^2C^{abe} C^{ecd}
\tilde{b}^{(4)}_{klmn}\int_{-\infty}^{\infty}
\frac{dq_z' dq_z''}{q_z' q_z''(q_z'+q_z''+p_z)}
\xi^b_l(q_z')\xi^c_m(-q_z'-q_z''-p_z)\xi^d_n(q_z'')=0.
\label{eom2}
\end{eqnarray}
Using the property of $\xi^a_j(p_z)$ in eq.(\ref{propxi}),
we obtain the result of the integration as
\begin{equation}
\int_{-\infty}^\infty\frac{\xi^a_j(p_z)}{p_z+\lambda}dp_z
=i\pi\xi^a_j(-\lambda).
\label{int}
\end{equation}
Then eq.(\ref{eom2}) after the use of this result simplifies to
\begin{equation}
\{ -\xi^a_1(-p_z)\delta_{k1}-\xi^a_2(-p_z)\delta_{k2}
+\xi^a_3(-p_z)\delta_{k3}\}
+\frac{g^2\pi^2}{2p_z^2}\tilde{b}^{(4)}_{klmn}C^{abe} C^{ecd}
\,\{\xi^b_l(0)\xi^c_m(-p_z)
-\xi^b_l(-p_z)\xi^c_m(0)\}\,\xi^d_n(0) 
= 0.
\label{eom3}
\end{equation}
We then make the following {\it{ansatz}}:
\begin{eqnarray}
\xi^a_1(p_z)&=&\sum_{n=0}^{\infty}\alpha^{a}_n p_z^{2n}
e^{-(p_z-\Delta)^2/\Delta^2},\nonumber\\
\xi^a_2(p_z)&=&\sum_{n=0}^{\infty}\beta^{a}_n p_z^{2n}
e^{-(p_z-\Delta)^2/\Delta^2},\nonumber\\
\xi^a_3(p_z)&=&\sum_{n=0}^{\infty}\gamma^{a}_n p_z^{2n}
e^{-(p_z-\Delta)^2/\Delta^2},
\label{ansatz}
\end{eqnarray}
which are also consistent with the property in eq.(\ref{propxi}).
Here $\Delta$ is a scale of mass dimension one. Then eq.(\ref{eom3})
simplifies to following three equations:
\begin{eqnarray}
\alpha_n^{a}+\frac{g^2\pi^2}{e^2}C^{abe}C^{ecd}
\Big\{\Big(\beta_0^{b}\alpha_{n+1}^{c}
-\alpha_0^{c}\beta_{n+1}^{b}\Big)\beta_0^{d}
+\Big(\gamma_0^{b}\gamma_{n+1}^{c}
-\gamma_0^{c}\gamma_{n+1}^{b}\Big)\alpha_0^{d}\Big\}&=& 0,
\label{alpha}\\
\beta_n^{a}-\frac{g^2\pi^2}{e^2}C^{abe}C^{ecd}
\Big\{\Big(\alpha_0{b}\alpha_{n+1}^{c}
-\alpha_0^{c}\alpha_{n+1}^{b}\Big)\beta_0^{d}
+\Big(\gamma_0^{b}\beta_{n+1}^{c}
-\beta_0^{c}\gamma_{n+1}^{b}\Big)\gamma_0^{d}\Big\}&=& 0,
\label{beta}\\
\gamma_n^{a}+\frac{g^2\pi^2}{e^2}C^{abe}C^{ecd}
\Big\{\Big(\alpha_0^{b}\gamma_{n+1}^{c}
-\gamma_0^{c}\alpha_{n+1}^{b}\Big)\alpha_0^{d}
-\Big(\beta_0^{b}\beta_{n+1}^{c}
-\beta_0^{c}\beta_{n+1}^{b}\Big)\gamma_0^{d}\Big\}&=& 0.
\label{gamma}
\end{eqnarray}
We now solve these equations for $\alpha_n^{a}$, $\beta_n^{a}$ and
$\gamma_n^{a}$ in case of $SU(2)$ gauge group. First we assume  
that $\alpha^{a}_n=f_n\delta^{a1}$, $\beta^{a}_n=g_n\delta^{a2}$
and $\gamma^{a}_n=h_n\delta^{a3}$. Upon substitution into 
eq.(\ref{alpha}), (\ref{beta}) and (\ref{gamma}) we obtain the relations 
for $SU(2)$ as
\begin{equation}
f_n+\frac{g^2\pi^2}{e^2}(g_0^2f_{n+1}-f_0g_0g_{n+1})=0,
\label{fn}
\end{equation}
\begin{equation}
g_n-\frac{g^2\pi^2}{e^2}(h_0^2g_{n+1}-h_0g_0h_{n+1})=0,
\label{gn}
\end{equation}
and
\begin{equation}
h_n+\frac{g^2\pi^2}{e^2}(f_0^2h_{n+1}-f_0h_0f_{n+1})=0.
\label{hn}
\end{equation}
From eq.(\ref{fn}) and (\ref{hn})
we get 
\begin{eqnarray}
g_{n+1} &=& \frac{g_0}{f_0}f_{n+1}+\frac{e^2}{g^2\pi^2}\frac{f_n}{f_0g_0},
\label{gnplus1}\\
h_{n+1} &=& \frac{h_0}{f_0}f_{n+1}+\frac{e^2}{g^2\pi^2}
\frac{h_n}{f_0^2}.
\label{hnplus1}
\end{eqnarray}
Substituting these two relations in eq.(\ref{gn}) we obtain
\begin{equation}
g_n=\frac{h_0^2}{f_0g_0}f_{n}+\frac{g_0h_0}{f_0^2}h_{n},
\label{gn1}
\end{equation}
which for $n=0$ reduces to 
\begin{equation}
\frac{g_0^2}{h_0^2}=1+\frac{g_0^2}{f_0^2}.
\label{cond1}
\end{equation}
Using eq.(\ref{gn1}) in eq.(\ref{fn}) we get
\begin{equation}
f_n+\frac{g^2\pi^2}{e^2}
\Big[(g_0^2-h_0^2)f_{n+1}-\frac{g_0^2h_0}{f_0}h_{n+1}\Big]=0.
\label{fn1}
\end{equation}
We compare eq.(\ref{hn}) and eq.(\ref{fn1}) and 
obtain after using eq.(\ref{cond1}) the relation
\begin{equation}
f_n=-\frac{g_0^2h_0}{f_0^3}h_n.
\label{fn2}
\end{equation}
Upon substitution of the value of $h_n$ into eq.(\ref{gn1}) 
we get
\begin{equation}
g_n=\frac{(h_0^2-f_0^2)}{f_0g_0}f_n,
\label{gn2}
\end{equation}
which for $n=0$ becomes
\begin{equation}
h_0^2=f_0^2+g_0^2,
\label{cond2}
\end{equation}
From eq.(\ref{cond1}) and eq.(\ref{cond2}) we get
\begin{equation}
h_0^2=g_0f_0.
\label{cond3}
\end{equation}
Then the use of eq.(\ref{cond1}) and (\ref{cond2}) in eq.(\ref{fn2}) 
and (\ref{gn2}) we obtain
\begin{equation}
\frac{f_n}{f_0}=\frac{g_n}{g_0}=\frac{h_n}{h_0}.
\label{fngnhn}
\end{equation}
It implies that $f_n$, $g_n$ and $h_n$ are linearly dependent 
coefficients of the expansion. 
If $x=\frac{f_0}{h_0}$ and $y=\frac{g_0}{h_0}$,
eq(\ref{cond2}) and (\ref{cond3}) become $x^2+y^2=1$ and $xy=1$. 
The solutions are $x=e^{\pm i\pi/6}$ and $y=x^{*}$.
Assume that $h_n=\delta_{n,0}\phi$, where $\phi$ is a 
constant, the solution reads
\begin{equation}
\bar{X}^a_{js}(p)=(2\pi)^4\phi\delta^{(3)}(p_\perp)\delta^{aj}
e^{is\theta_j}
e^{-(p_z-\Delta)^2/\Delta^2},
\label{Xbar}
\end{equation}
where,
\begin{equation}
s=\pm,\,\,\theta_j= \frac{\pi}{6}(\delta^{j1}-\delta^{j2}).
\end{equation}

\section{Stability of the solution}

To address the issue of the stability of the obtained 
solution we compute the vacuum energy density $\epsilon(\phi)$
of the system. The solutions will be stable if $\epsilon(\phi)$ 
has a minimum with respect to $\phi$ and it remains 
bounded even for limited fluctuation of $\phi$ around 
the minimum. From trace anomaly\cite{Peskin} we know
that the trace of the energy momentum tensor
\begin{equation}
\Theta^\mu_\mu = \frac{\beta(g)}{2 g^3}G^a_{\alpha\beta}G^{a\alpha\beta},
\label{traceanomaly}
\end{equation}
where $\beta(g)$ ($=\mu\frac{\partial g}{\partial\mu}$)
is negative for pure NAGT. Lorentz invariance
requires that $\langle 0\mid\Theta_{\mu\nu}\mid 0\rangle
=\epsilon g_{\mu\nu}$ which is also consistent with the fact that
$\Theta_{00}=\epsilon$. We therefore obtain  
$\langle 0\mid\Theta_{\mu\nu}\mid 0\rangle
=4\epsilon$ and then the use of eq.(\ref{traceanomaly}) gives
\begin{equation}
\epsilon= \frac{\beta(g)}{8 g^3}\langle 0\mid
G^a_{\alpha\beta}G^{a\alpha\beta}\mid 0\rangle.
\label{energydensity}
\end{equation}
In the following we 
first expand the action around the solutions 
to quadratic order in fluctuations and then 
proceed to compute $\langle 0\mid
G^a_{\alpha\beta}G^{a\alpha\beta}\mid 0\rangle$
using this expanded action around the solutions.
Solutions in eq.(\ref{ansatz}) are dependent on a mass scale 
$\Delta$ which is dynamic in nature. It will be shown in the later 
part of this section that $\Delta$ remains less than $1$ when 
$g\gg 1$. As we are interested to check the stability of the 
solution in the strong coupling limit, we take 
$\Delta\ll 1$ and expand eq.(\ref{Xbar}) in the following in powers of 
$\Delta$. The regularized solution then reads, 
\begin{equation}
\bar{X}^a_{js}(p) = (2\pi)^4
\sqrt{\pi}\phi\Delta\delta^{(3)}(p_\perp)
\delta(p_z-\Delta-\zeta)\delta^{aj} e^{is\theta_j},
\end{equation}
where we take $\zeta\rightarrow 0$ at the end of our calculation.
We introduce $3(N_c^2-1)$ sources and write the partition function 
of eq.(\ref{Zk}) as
\begin{equation}
Z[J] = \int\mathcal{D}G\delta[I[G]]
e^{iS[G,J]},
\label{ZJ}
\end{equation}
where
\begin{equation}
S[G,J] = \int\frac{d^4k}{(2\pi)^4} 
\{-\frac{1}{4}G^a_{\mu\nu}(k)G^{a\mu\nu}(-k) 
+ X^a_j(k)J^a_j(-k)\}.
\label{SGJ}
\end{equation}
Consider an operator $O({\mathbf E}^a, {\mathbf B}^a)$
which is a function of non abelian electric and magnetic
fields. The vacuum expectation value
\begin{equation}
\langle 0\mid O({\mathbf E}^a, {\mathbf B}^a)\mid 0\rangle
= \frac{1}{Z[0]} 
\int\mathcal{D}G\delta[I[G]]\,
O({\mathbf E}^a, {\mathbf B}^a)
 e^{iS[G, J]}
\end{equation}
Then, after performing the integration over $E^a_x$, $E^a_y$ and $B^a_z$
as discussed in section II we obtain
\begin{equation}
\langle 0\mid O({\mathbf E}^a, {\mathbf B}^a)\mid 0\rangle
=\frac{{\cal N}}{Z[0]}\int{\cal{D}}X \tilde{O}(X^a_j)  
e^{iS [X,J]}
=\frac{1}{Z[0]}\tilde{O}\left
(\frac{(2\pi)^4}{i}\frac{\delta}{\delta J}\right)
Z[J]\mid_{J=0},
\label{aveO}
\end{equation}
where the action of eq.(\ref{SX}) is modified as
\begin{equation}
S[X,J]=S[X]+\int\frac{d^4p}{(2\pi)^4}X^a_j(p)J^a_j(-p).
\label{SXJ}
\end{equation}
Moreover, according to eq.(\ref{f1}), (\ref{f2}) and (\ref{f3})  
$f^a_j$ are functions of $X^a_j$ ($j=1,2,3$), we use
\begin{equation}
\tilde{O}(X^a_j)=O(f^a_j,X^a_j).
\end{equation}
To compute the vacuum average of the operator we 
adopt the stationary phase approximation in $Z[J]$:
\begin{equation}
Z[J]={\cal N}\int{\cal D}X e^{iS[X,J]}.
\label{ZJ1}
\end{equation}
We expand the action about the classical solution and 
keep terms to order $(X-\bar{X})^2$:
\begin{equation}
S[X,J] = \sum_{s=\pm}S[\bar{X}_s]
+\frac{1}{2}\sum_{s=\pm}
\int \frac{d^4p}{(2\pi)^4}
\Phi^a_{js}(p)({\cal{M}}_s^{-1})^{(ab)}_{jk}(p))\Phi^b_{ks}(-p)
+\int \frac{d^4p}{(2\pi)^4}J^a_k(p)X^a_k(-p)+0(\Phi^3)
\label{SXJs}
\end{equation}
where $\Phi^a_{js)a}(p)= X^a_j(p)-\bar{X}^a_{js}(p)$.
$\Phi^a_{js}(p)$ is assumed as  
a very slowly varying function of $p_z$ over the momentum 
scale $\Delta$ ($\Delta\ll 1$) and
\begin{equation}
({\cal{M}}_s^{-1})^{(ab)}_{jk}(p)
=\delta^{ab}(M^{-1})_{jk}(p)
+\frac{\bar{\phi}}{p_z(p_z+\Delta)}C^{abc}e^{is\theta_c}
p_\perp\lambda^{(\perp)}_{cjk}
-\frac{\bar{\phi}^2}{p_z(p_z+2\Delta)}
C^{ace}C^{ebd} e^{is(\theta_c+\theta_d)}\alpha^{(cd)}_{jk}.
\label{calMsinv}
\end{equation}
We have defined
\begin{equation}
\bar{\phi}=\sqrt{\pi}g\phi,
\end{equation}
where the non-vanishing components
of $\lambda$s and $\alpha$s are as follows:
\begin{eqnarray}
\lambda^{(0)}_{ajk}
&=&\delta^{a1}(\delta_{j1}\delta_{k3}+\delta_{j3}\delta_{k1})
+\delta^{a2}(\delta_{j2}\delta_{k3}+\delta_{j3}\delta_{k2})
-2\delta^{a3}(\delta_{j1}\delta_{k1}+\delta_{j2}\delta_{k2})\nonumber\\
\lambda^{(1)}_{ajk}
&=&\delta^{a1}(\delta_{j1}\delta_{k2}+\delta_{j2}\delta_{k1})
+2\delta^{a2}(\delta_{j3}\delta_{k3}-\delta_{j1}\delta_{k1})
-\delta^{a3}(\delta_{j2}\delta_{k3}+\delta_{j3}\delta_{k2})\nonumber\\
\lambda^{(2)}_{ajk}
&=&2\delta^{a1}(\delta_{j2}\delta_{k2}-\delta_{j3}\delta_{k3})
-\delta^{a2}(\delta_{j2}\delta_{k1}+\delta_{j1}\delta_{k2})
+\delta^{a3}(\delta_{j1}\delta_{k3}+\delta_{j3}\delta_{k1})
\end{eqnarray}
and
\begin{eqnarray}
& &\alpha^{(11)}_{jk}
=\delta_{j3}\delta_{k3}-\delta_{j2}\delta_{k2};
~~\alpha^{(12)}_{jk}=2\delta_{j2}\delta_{k1}-\delta_{j1}\delta_{k2};
~~\alpha^{(13)}_{jk}=\delta_{j1}\delta_{k3}-2\delta_{j3}\delta_{k1};
\nonumber\\
& &\alpha^{(21)}_{jk}
=2\delta_{j1}\delta_{k2}-\delta_{j2}\delta_{k1};
~~\alpha^{(22)}_{jk}=\delta_{j3}\delta_{k3}-\delta_{j1}\delta_{k1};
~~\alpha^{(23)}_{jk}=\delta_{j2}\delta_{k3}-2\delta_{j3}\delta_{k2};
\nonumber\\
& &\alpha^{(31)}_{jk}
=\delta_{j3}\delta_{k1}-2\delta_{j1}\delta_{k3};
~~\alpha^{(32)}_{jk}=\delta_{j3}\delta_{k2}-2\delta_{j2}\delta_{k3};
~~\alpha^{(33)}_{jk}=\delta_{j1}\delta_{k1}+\delta_{j2}\delta_{k2}.
\end{eqnarray}

To get rid of the linear terms of $X(p)$ in eq.(\ref{SXJs}), we set
$X(p)=X^\prime(p)+C(p)$ and choose 
\begin{equation}
C(p) =-\frac{1}{2}({\rm Re}Q_+(-p))J(p)-({\rm Re}Q_+(-p))
{\rm Re}\{Q_+^{-1}(-p)\bar{X}_+(p)\},
\end{equation}
where
\begin{equation}
(Q^{-1}_s)^{(ab)}_{jk}(p)=\frac{1}{2}\{({\cal{M}}_s^{-1})^{(ab)}_{jk}(p)
+({\cal{M}}_s^{-1})^{(ba)}_{kj}(-p)\}
\label{Qsinv}
\end{equation}
Then eq.(\ref{SXJs}) in terms of $X^\prime(p)$ reads
\begin{eqnarray}
S[X,J]&=& 2 {\rm Re}S[\bar{X}_+]
+\int \frac{d^4p}{(2\pi)^4}
{\rm Re}\{\bar{X}_+(p)Q^{-1}_+(p)\bar{X}_+(-p)\}\nonumber\\
& &-\int \frac{d^4p}{(2\pi)^4}
{\rm Re}\{Q^{-1}_+(-p)\bar{X}_+(p)\}({\rm Re}Q_+(p))
{\rm Re}\{Q^{-1}_+(p)\bar{X}_+(-p)\} \nonumber\\
& &-\frac{1}{4}\int\frac{d^4p}{(2\pi)^4}J(p)({\rm Re}Q_+(p))J(-p)
-\int\frac{d^4p}{(2\pi)^4}
{\rm Re}\{Q^{-1}_+(-p)\bar{X}_+(p)\}({\rm Re}Q_+(p))J(-p)\nonumber\\
& &+\int\frac{d^4p}{(2\pi)^4}X^\prime(p)({\rm Re}Q^{-1}_+(p))X^\prime(-p).
\end{eqnarray}
Then we change the integration variable in eq.(\ref{ZJ1})
from $X$ to $X'$ and integrate over $X'$ to obtain
\begin{equation}
Z[J] = \frac{{\cal N}}
{\sqrt{{\rm Det}({\rm Re}Q_+^{-1})}}
\exp\Big[iS_0
-\frac{i}{4}\int \frac{d^4p}{(2\pi)^4}J(p)({\rm Re}Q_+(p))J(-p)
-i\Delta\eta_k^b J_k^b(0,-\Delta-\zeta)
\Big],
\label{ZJfinal}
\end{equation}
where, 
\begin{eqnarray}
S_0 &=& 2 {\rm Re}S[\bar{X}_+]
-\int \frac{d^4p}{(2\pi)^4}
{\rm Re}\{Q^{-1}_+(-p)\bar{X}_+(p)\}({\rm Re}Q_+(p))
{\rm Re}\{Q^{-1}_+(p)\bar{X}_+(-p)\}\nonumber\\
& &+\int\frac{d^4p}{(2\pi)^4}
{\rm Re}\{\bar{X}_+(p)Q^{-1}_+(p)\bar{X}_+(-p)\},\\
\eta^b_k&=&\frac{\bar{\phi}}{g}
\Big\{
(M^{-1})_{aj}\cos\theta_a+\frac{\bar{\phi}^2}{3\Delta^2}
(\alpha^{(cc)}-\alpha^{(ac)})_{cj}\cos(\theta_a+2\theta_c)
\Big\}\nonumber\\
&\times&
\Big[
\delta^{ab}M_{jk}-\frac{\bar{\phi}^2}{3\Delta^2}
\{
\delta^{ab}\cos 2\theta_d(M_{jm}\alpha^{(dd)}_{mm'}M_{m'k})
-\cos(\theta_a+\theta_b)(M_{jm}\alpha^{(ba)}_{mm'}M_{m'k})
\}\nonumber\\
& &+\frac{\bar{\phi}^4}{9\Delta^4}\cos 2\theta_d
\{
\delta^{ab}\cos2\theta_f
(M_{jm}\alpha^{(ff)}_{mn}M_{nn'}\alpha^{(dd)}_{n'm'}M_{m'k})
-\cos(\theta_a+\theta_b)
(M_{jm}\alpha^{(ba)}_{mn}M_{nn'}\alpha^{(dd)}_{n'm'}M_{m'k})
\}
\nonumber\\
& &-\frac{\bar{\phi}^4}{9\Delta^4}\cos(\theta_b+2\theta_d)
\{
\delta^{ad}\cos2\theta_f
(M_{jm}\alpha^{(ff)}_{mn}M_{nn'}\alpha^{(bd)}_{n'm'}M_{m'k})
-\cos(\theta_a+\theta_d)
(M_{jm}\alpha^{(da)}_{mn}M_{nn'}\alpha^{(bd)}_{n'm'}M_{m'k})
\}
\Big]
\end{eqnarray}
and the convention $J(p)=J(p_\perp,p_z)$ has been adopted.

We take the gauge field operator
\begin{equation} 
O= G^{a\mu\nu}(0)G^a_{\mu\nu}(0)
=\int \frac{d^4p}{(2\pi)^4}
\frac{d^4q}{(2\pi)^4}G^{a\mu\nu}(p)G^a_{\mu\nu}(q)
\end{equation}
and the use of eq.(\ref{aveO}) gives the vacuum average as 
\begin{equation}
\langle 0\mid G^{a\mu\nu}(0)G^a_{\mu\nu}(0)\mid 0\rangle
= O_2 + O_3 + O_4.
\end{equation}
The expressions for $O_2$, $O_3$ and $O_4$ are as follows:
\begin{eqnarray}
O_2 &=&
2(2\pi)^8\Delta^2(\eta^a_1\eta^a_1
+\eta^a_2\eta^a_2-\eta^a_3\eta^a_3)
-2i\int \frac{d^4p}{(2\pi)^4}
(M^{-1})_{jk}(p)({\rm Re}Q_{+})^{(aa)}_{kj}(p),
\label{O2}\\
O_3 &=&
4igC^{abc}\int\frac{d^4p}{(2\pi)^4}
\frac{b_{jkl}^{(3)}(p)}{p_z^2}
\Big[\eta_{k}^b({\rm Re}Q_{+})^{(ca)}_{lj}(p)
+\eta_{l}^c({\rm Re}Q_{+})^{(ba)}_{kj}(p)\Big],
\label{O3}\\
O_4 &=& 2g^2C^{ab_1c_1}C^{ab_2c_2}b_{jklm}^{(4)}
\Big[-\eta^{b_1}_{j}\eta^{c_1}_{k}
\eta^{b_2}_{l}\eta^{c_2}_{m}
-i\int\frac{d^4p}{(2\pi)^4}\frac{1}{p_z^2}\Big\{
({\rm Re}Q_{+})^{(c_2 b_1)}_{mj}(p)
\eta^{b_2}_{l}\eta^{c_1}_{k}
+({\rm Re}Q_{+})^{(c_2 c_1)}_{mk}(p)
\eta^{b_1}_{j}\eta^{b_2}_{l}\nonumber\\
& &+({\rm Re}Q_{+})^{(c_2 b_2)}_{ml}(p)
\eta^{b_1}_{j}\eta^{c_1}_{k}
+({\rm Re}Q_{+})^{(b_2 b_1)}_{lj}(p)
\eta^{c_2}_{m}\eta^{c_1}_{k}
+({\rm Re}Q_{+})^{(b_2c_1)}_{lk}(p)
\eta^{b_1}_{j}\eta^{c_2}_{m}
+({\rm Re}Q_{+})^{(c_1b_1)}_{kj}(p)
\eta^{b_2}_{l}\eta^{c_2}_{m} \Big\}\nonumber\\
& &+\int \frac{d^4p}{(2\pi)^4}
\frac{d^4q}{(2\pi)^4}
\frac{1}{p_z^2 q_z^2}
\Big\{
({\rm Re}Q_{+})^{(b_2 b_1)}_{lj}(p)({\rm Re}Q_{+})^{(c_2 c_1)}_{mk}(q)
+({\rm Re}Q_{+})^{(b_2 c_1)}_{lk}(p)({\rm Re}Q_{+})^{(c_2 b_1)}_{mj}(q)
\nonumber\\
& &+({\rm Re}Q_{+})^{(c_1 b_1)}_{kj}(p)
({\rm Re}Q_{+})^{(c_2 b_2)}_{ml}(q)\Big\}\Big].
\label{O4}
\end{eqnarray}

We evaluate the momentum integrals in $O_2$, $O_3$ and $O_4$
for $SU(2)$ gauge group.
First we change the momentum integrals from Minkowski to Euclidean
space: $p_4=ip_0$ and $d^4p=-id^4p_E$ where the Euclidean four
momenta $p_E=(p_4,{\mathbf p})$. 
We then use the system of coordinates
in a manner such that the triad ($p_x$, $p_y$, $p_4$) form a spherical
system of coordinates and write
$p_4=r\cos\theta$, $p_x=r\sin\theta\cos\theta_1$
and $p_y=r\sin\theta\sin\theta_1$. 
The integrals are evaluated over the following ranges:  
$0\le r\le\Lambda$, $0\le\theta\le\pi$ and $0\le\theta_1\le 2\pi$,
where $\Lambda$ is an ultraviolet cut off used in the calculation. 
Since the integrands are found to be even function of $p_z$ we 
fold the region of integration from $-\Lambda\le p_z\le\Lambda$
to $\Lambda_{ir}\le p_z\le\Lambda$, where the small and positive 
quantity $\Lambda_{ir}$ has been used as an infrared cut off  
in the calculation. 
We make the integration variables dimensionless:
$x= r/\Lambda$ and $y=p_z/\Lambda$ so that $0\le x\le 1$ and 
$\frac{\Lambda_{ir}}{\Lambda}\le y\le 1$. 
The form of the inverse of ${\rm Re}Q_+^{-1}$, which appears 
as factrors in the integrands of the vacuum average of the 
operator $O$, are given in the appendix. 
Assuming $\Delta<\Lambda_{ir}$ we expand $\beta^{(I)(ab)}_{jk}(p)$
in powers of $\Delta^2/p_z^2$ and evaluate the integrals.
The obtained vacuum average of $O$ is ultraviolet divergent when
$\Lambda$ goes to $\infty$ for $\Lambda_{ir}\ne 0$. 
We use minimal subtraction scheme to obtain the renormalized 
vacuum average and then take $\Lambda_{ir}$ going to zero limit.
Thus we obtain the renormalized vacuum average as
\begin{eqnarray}
\langle 0\mid G^{a\mu\nu}(0)G^a_{\mu\nu}(0)\mid 0\rangle
&=&\Delta^4\bigg[
\left(\frac{\phi}{\Delta}\right)^4\Big\{
q^\prime_2g^4+g^4\sum_{n=0}^{n=3}  c_n \left(g\frac{\phi}{\Delta}\right)^{2n}
+\frac{1}{g^2}\sum_{n=0}^{n=12} d_n\left(g\frac{\phi}{\Delta}\right)^{2n}
+g^2\sum_{n=0}^{n=12}d^\prime_n\left(g\frac{\phi}{\Delta}\right)^{2n}
+d^{\prime\prime}_0 g^6\Big\}\nonumber\\
& &+\left(\frac{\phi}{\Delta}\right)^2
\Big\{\sum_{n=0}^{n=6}q_n\left(g\frac{\phi}{\Delta}\right)^{2n}
+q^\prime_3+d^{\prime\prime}_1g^4\Big\}\bigg]\nonumber\\
&=&\Delta^4\sum_{n=1}^{n=14} B_n(g)\left(\frac{\phi}{\Delta}\right)^{2n},
\label{Gsquare}
\end{eqnarray}
where
\begin{eqnarray}
B_1(g) &=& q_0+q^\prime_3+d^{\prime\prime}_1g^4, \,\,\, 
B_2(g) = q_1g^2+c_0g^4+\frac{d_0}{g^2}+d^\prime_0g^2
+d^{\prime\prime}_0g^6+q^\prime_2g^4,\\ 
B_3(g) &=& q_2g^4+c_1g^6+d_1+d^\prime_1g^4,\,\,\,
B_4(g) = q_3g^6+c_2g^8+d_2g^2+d^\prime_2g^6,\\
B_5(g) &=& q_4g^8+c_3g^{10}+d_3g^4+d^\prime_3g^3,\,\,\,
B_6(g) = q_5g^{10}+d_4g^6+d^\prime_4g^{10},\,\,\,
B_7(g) = q_6g^{12}+d_5g^8+d^\prime_5g^{12}
\end{eqnarray}
and
\begin{equation}
B_n(g)=d_{n-2}g^{(2n-6)}+d^\prime_{n-2}g^{(2n-2)}
\end{equation}
for $8\le\,n\,\le 14$.
The numerical values of the coefficients are listed in 
Table\ref{tableO2}, \ref{tableO3}, \ref{tableO4one} and \ref{tableO4two}.
\begin{table}[!h]
{\begin{tabular}{|c|c|c|c|c|c|c|}
\hline
$q_0$  & $q_1$ & $q_2$ & $q_3$ & $q_4$ & $q_5$ & $q_6$\\ 
\hline
$2.73\times 10^{12}$ & $6.012\times 10^{11}$ & $-3.397\times 10^{11}$ & $8.2513\times 10^{9}$
& $-1.07\times 10^{10}$ & $-2.858\times 10^{8}$ & $2.1785\times 10^{8}$\\ 
\hline
$q_2^\prime$ & $q_3^\prime$ \\
\cline{1-2}
$-1.44$ & $-63.66$\\
\cline{1-2}
\end{tabular}}
\caption{Coefficients of $O_2$}
\label{tableO2}
\end{table}
\begin{table}[!h]
{\begin{tabular}{|c|c|c|c|}
\hline
$c_0$  & $c_1$ & $c_2$ & $c_3$\\
\hline
$-79.2303$ & $1.8284\times 10^{2}$ & $-12.42$ & $5.175$\\
\hline
\end{tabular}}
\caption{Coefficients of $O_3$}
\label{tableO3}
\end{table}
\begin{table}[!h]
{\begin{tabular}{|c|c|c|c|c|c|c|}
\hline
$d_0$  & $d_1$ & $d_2$ & $d_3$ & $d_4$ & $d_5$ & $d_6$\\
\hline
$-5.80\times 10^{11}$ & $3.50\times 10^{12}$
& $-1.7\times 10^{12}$ & $3.5\times 10^{11}$
& $-1.15\times 10^{11}$ & $2.52\times 10^{10}$
& $-2.83\times 10^{9}$\\
\hline
$d_7$ & $d_8$ & $d_9$ & $d_{10}$ & $d_{11}$ & $d_{12}$\\
\cline{1-6}
$6.20\times 10^{8}$ & $-9.16\times 10^{7}$
& $-6.884\times 10^{6}$ & $6.34\times 10^{5}$
& $-6.101\times 10^{3}$ & $-8.87\times 10^{2}$\\
\cline{1-6}
\end{tabular}}
\caption{Coefficients of the non-integral terms of $O_4$}
\label{tableO4one}
\end{table}
\begin{table}[!h]
{\begin{tabular}{|c|c|c|c|c|c|c|}
\hline
$d^\prime_0$ & $d^\prime_1$ & $d^\prime_2$ & $d^\prime_3$
& $d^\prime_4$ & $d^\prime_5$ & $d^\prime_6$\\
\hline 
$-2.03\times 10^{12}$
& $3.15\times 10^{11}$
& $-1.41\times 10^{10}$
& $2.41\times 10^{10}$
& $-8.84\times 10^{12}$
& $4.48\times 10^{25}$
& $-4.51\times 10^{25}$\\
\hline
$d^\prime_7$ & $d^\prime_8$ & $d^\prime_9$ & $d^\prime_{10}$
& $d^\prime_{11}$ & $d^{\prime}_{12}$ & $d^{\prime\prime}_0$\\
\hline
$2.12\times 10^{25}$
& $-5.08\times 10^{24}$
& $8.90\times 10^{23}$
& $-1.61\times 10^{23}$
& $9.78\times 10^{21}$
& $-1.56\times 10^{21}$
& $-9.96\times 10^{13}$\\
\hline
$d''_1$\\
\cline{1-1}  
$1.20\times 10^{-5}$\\
\cline{1-1}
\end{tabular}}
\caption{Coefficients of the single and double integral terms of 
$O_4$.}
\label{tableO4two}
\end{table}

We use eq.(\ref{energydensity}) to obtain the energy density as 
\begin{equation}
\epsilon(\phi)= -\frac{\mid\beta(g)\mid}{8g^3}
\Delta^4\sum_{n=1}^{n=14} B_n(g)\left(\frac{\phi}{\Delta}\right)^{2n},
\label{ephi}
\end{equation}
\begin{figure}[!th]
\begin{center}
\includegraphics[scale=0.4]{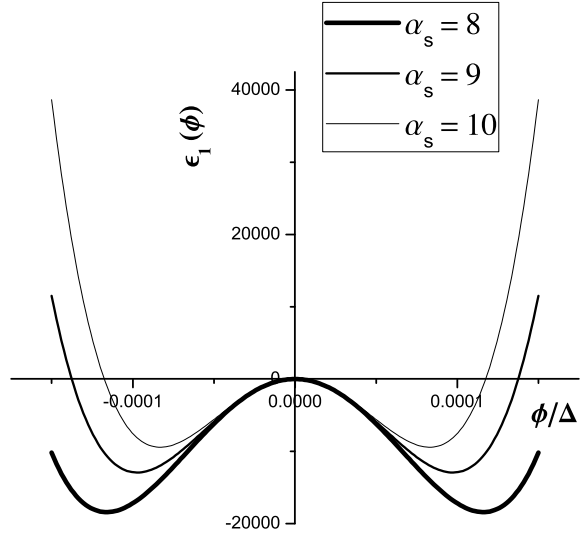}
\end{center}
\vskip -0.45in
\caption[]{(Color online) Plot of $\epsilon_1(\phi)$ versus
$\phi/\Delta$ in the non-perturbative regime.} 
\label{fige1lg}
\end{figure}
which is an even polynomial function of $\phi$ of degree $28$.
Apart from having a maximum at $\phi=0$, it must have at least one 
pair of minima at non-zero $\phi$ depending on the the strength of strong 
coupling constant $\alpha_s$ ($=g^2/4\pi$). We have studied the behaviour of
$\epsilon(\phi)$ and its first pair of minima in the strong coupling 
region, where $\alpha_s > 1$.
Some of our results are plotted in Fig.\ref{fige1lg}, where
the plots of $\epsilon_1(\phi)=\frac{8g^3\epsilon(\phi)}
{\mid\beta(g)\mid\Delta^4}$ versus $\phi/\Delta$ for 
different $\alpha_s$ are shown. 
Our numerical evaluation shows that the first pair of minima
behaves as $\phi_0=\pm b_1\Delta/\alpha_s^{a_1}$ in the 
strong coupling region, where
$a_1=1.50165 \pm 0.003$, $b_1=0.00261 \pm 3.12944\times 10^{-6}$.

Since, $\epsilon(\phi)$ is a bounded function under a small fluctuations 
of $\phi$ around the minima, obtained solution remains stable when  
the coupling is high.
According to eq.(\ref{Gsquare}) the gluon condensate at the 
minima of the energy density curve in the strong coupling region reads
\begin{equation}
\langle\alpha_s G^2\rangle_0 = \Delta^4
\sum_{n=1}^{14} B_n(g) b_1^{2n}\alpha_s^{-2na_1+1}.
\end{equation}
\begin{figure}[!th]
\begin{center}
\includegraphics[scale=0.4]{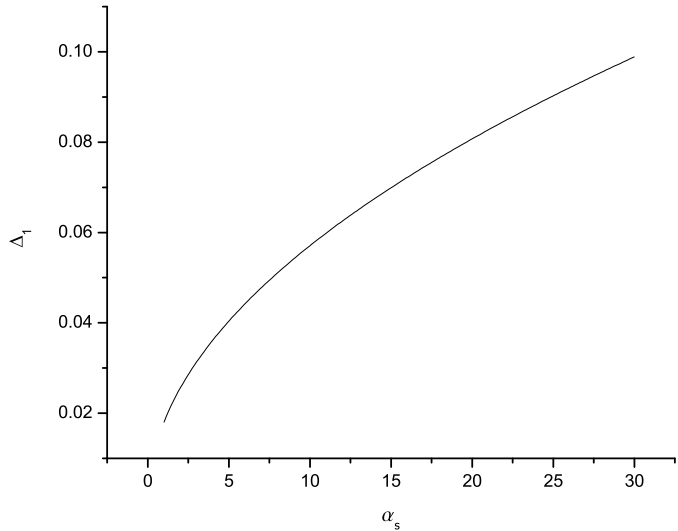}
\end{center}
\vskip -0.45in
\caption[]{(Color online) Plot of $\Delta_1$ versus
$\alpha_s$ in the strong coupling regime.} 
\label{figD1lg}
\end{figure}
The plot of $\Delta_1$ 
($=\Delta/\sqrt[4]{\langle\alpha_s G^2\rangle_0}$) versus
$\alpha_s$ is shown in the Fig.\ref{figD1lg}, where 
$\Delta_1$ is a monotonically increasing function of $\alpha_s$
and it remains less than $1$ for quite a large value of 
$\alpha_s$ in the region $1\le\alpha_s\le 3069$. Since,
for $SU(2)$ $\langle\alpha_s G^2\rangle_0=0.0314 {\rm GeV}^4$
\cite{Furnstahl1990}, $\Delta$ remains less than $1$ in this 
region of $\alpha_s$. It justifies the validity of smallness 
of $\Delta$ in a region of strong coupling, where 
$1\le\alpha_s < 3069$.

\section{Conclusion}

We have taken the gauge fixed partition function of pure NAGT in axial gauge
using the field strength formalism as given by Halpern.
We have integrated out the variables $E_x^a$, $E_y^a$ and $B_z^a$
using Bianchi Identities in momentum space
and have obtained an effective action of the remaining $3(N_c^2-1)$
variables $E_z^a$, $B_x^a$ and $B_y^a$.
We have obtained a static solution of EOM of the effective theory
in terms of parameters $\phi$ and $\Delta$.
The solutions in the momentum space exhibit a Gaussian nature in
the $z$ component of momentum and are proportional to Dirac delta
functions of the remaining components of momentum.
To check the stability of the solutions first
we expand the effective action around the
solutions to quadratic order in fluctuations and
then compute the vacuum average
$\langle 0\mid G^{a\mu\nu}(0)G^a_{\mu\nu}(0)\mid 0 \rangle$
using this expanded action around the solutions.
Then we use trace anomaly to obtain the energy density as an
even polynomial function of $\phi$ of degree $28$. 

Apart from having a maximum at $\phi=0$, energy density
has several maxima and minima at non-zero $\phi$, which are 
symmetrically situated around $\phi=0$. The energy density 
remains stable under fluctuations around any one of the 
minima and hence ensures the stability of the obtained solution.
The negative energy density at the minimum indicates that the 
system itself is bound. The behaviour of first pair of minima 
at $\phi=\pm\phi_0$ close to $\phi=0$ has been studied in the 
strong coupling region, where $\phi_0$ falls with the 
increase of $\alpha_s$ following power law. 
The parameter $\Delta$ is found to be proportional 
to $\sqrt[4]{\langle\alpha_s G^2\rangle_0}$, where the constant
of proportionality increases monotonically with $\alpha_s$.
The smallness of $\Delta$ in the region $1\le\alpha_s <3069$
indicates the validity of our approximation scheme in the 
strong coupling region.


\appendix
\section{Real part of $Q_+(p)$}
\label{appendixA}

Given ${\cal M}_s^{-1}$ in eq.(\ref{calMsinv}), we compute 
the real part of $Q_+^{-1}$ using eq.(\ref{Qsinv}) and then 
obtain its inverse which is as follows: 
\begin{equation}
({\rm Re}Q_+)^{(ab)}_{jk}(p)=\delta^{ab}M_{jk}(p)
+\sum_{I=1}^{\infty}\beta^{(I)(ab)}_{jk}(p)\bar{\phi}^I
\end{equation}
Some of the coefficients for non-zero $I$ in case of $SU(2)$
are as follows:
\begin{eqnarray}
\beta^{(1)(ab)}_{jk}(p) &=&-\frac{\epsilon^{abc}}{(p_z^2-\Delta^2)}
\cos\theta_c\{ M(p_{\perp}\lambda^{(\perp)}_c)M\}_{jk},\\
\beta^{(2)(ab)}_{jk}(p) &=&
-\frac{\epsilon^{cbd}}{(p_z^2-\Delta^2)}\beta^{(1)(ac)}_{jl}(p)
\{ (p_{\perp}\lambda^{(\perp)}_d)M\}_{lk}\cos\theta_d
+\frac{\epsilon^{ade}\epsilon^{ebf}}{(p_z^2-4\Delta^2)}
M_{jl}\alpha^{(df)}_{ln}M_{nk}\cos(\theta_d+\theta_f),\\
\beta^{(3)(ab)}_{jk}(p) &=&
-\frac{\epsilon^{cbd}}{(p_z^2-\Delta^2)}\beta^{(2)(ac)}_{jl}(p)
\{ (p_{\perp}\lambda^{(\perp)}_d)M\}_{lk}\cos\theta_d
+\frac{\epsilon^{cde}\epsilon^{ebf}}{(p_z^2-4\Delta^2)}
\beta^{(1)(ac)}_{jl}(p)\alpha^{(df)}_{ln}M_{nk}\cos(\theta_d+\theta_f),\\
\beta^{(4)(ab)}_{jk}(p) &=&
-\frac{\epsilon^{cbd}}{(p_z^2-\Delta^2)}\beta^{(3)(ac)}_{jl}(p)
\{ (p_{\perp}\lambda^{(\perp)}_d)M\}_{lk}\cos\theta_d
+\frac{\epsilon^{cde}\epsilon^{ebf}}{(p_z^2-4\Delta^2)}
\beta^{(2)(ac)}_{jl}(p)\alpha^{(df)}_{ln}M_{nk}\cos(\theta_d+\theta_f).
\end{eqnarray}

\end{document}